\documentclass[useAMS,usegraphicx,usenatbib]{mn2e}
\usepackage{txfonts}
\renewcommand\pi\upi

\date{
{\it Monthly Notices of the Royal Astronomical Society} {\bf 360}, 492-498 (2005)
\hfill doi:10.1111/j.1365-2966.2005.09078.x}
\pagerange{\pageref{start}--\pageref{finish}}
\pubyear{2005}
\volume{360}

\title[Hypervirial Models]
{Hypervirial Models of Stellar Systems}

\author[N.~W.~Evans \& J.~An]
{N.~W.~Evans\thanks{E-mail: nwe,jin@ast.cam.ac.uk} and
J.~An\footnotemark[1]\\
Institute of Astronomy, University of Cambridge,
Madingley Road, Cambridge CB3 0HA, UK}

\begin{document} 
\label{start}
\maketitle

\begin{abstract}
A family of cusped potential-density pairs is presented for modelling
galaxies and dark haloes.  The density profile is cusped like $\rho
\sim r^{p-2}$ at small radii. The distribution function is simple and
takes the form $f \propto L^{p-2} E^{(3p+1)/2}$ (where $E$ is the
binding energy and $L$ is the angular momentum). The models all
possess the remarkable property that the virial theorem holds locally,
from which they earn their name as the {\it hypervirial
family}. Famously, this property was first discovered by Eddington to
hold for the Plummer model in 1916. In fact, the seductive properties
of the Plummer model extend to the whole hypervirial family, including
the members possessing the cosmologically important cusps of $\rho
\sim r^{-1}$ or $\rho \sim r^{-3/2}$ or $\rho \sim r^{-4/3}$. The
intrinsic and projected properties of the family of models are
discussed in some detail.
\end{abstract}


\section{Introduction}

The allure of spherical models is their simplicity. Certainly, there
are few exactly spherical galaxies or bulges. However, spherical models
are still useful as representations of galactic nuclei, where the
kinematics are usually dominated by a central cusp \citep[see e.g.,][]
{De93,Tr94}. Almost nothing is firmly known concerning the
shapes of the dark matter haloes around disc galaxies.  So, it is very
reasonable to use spherical models to investigate, for example, the
kinematics of distant halo stars and satellite galaxies of the Milky
Way \citep{Wh85,KL92,WE99}.
Globular clusters with long relaxation times, such as $\omega$
Centauri, roundish dwarf spheroidal galaxies such as Draco
\citep[e.g.,][]{Wi02}, and clusters of galaxies are also
possible applications for spherical models.

\citet{Pl11} set the standard when he introduced the model with
potential $\psi$ and mass density $\rho$ as a fit to the outer parts
of globular clusters,
\begin{equation}
\psi = \frac{GM}{(a^2 + r^2)^{1/2}},
\qquad
\rho = \frac{3M}{4\pi}\frac{a^2}{(a^2+r^2)^{5/2}}.
\end{equation}
The real beauty of this model lies in two facts discovered -- not by
Plummer -- but by \citet{Ed16}. First, the isotropic Plummer model has a
very simple distribution function.  In fact, Eddington showed that the
isotropic distribution function of any spherical model is
available immediately as a quadrature \citep[p. 237]{Ed16,BT87},
\begin{equation}
f(E) = \frac{1}{\sqrt{8}\pi^2} \left(\int_0^E
\frac{\mathrm d^2\rho}{\mathrm d\psi^2}\frac{\mathrm d\psi}{\sqrt{E-\psi}}
+\frac{1}{\sqrt{E}}\left.\frac{\mathrm d\rho}{\mathrm d\psi} \right|_{\psi=0}
\right).
\label{eq:eddington}
\end{equation}
Here, $f$ is the distribution function, which depends on the specific
binding energy $E$ of the stars alone. Substituting the particular case of
the Plummer model, Eddington found the very simple answer
\begin{equation}
f(E) \propto E^{7/2}.
\label{eq:plummerdf}
\end{equation}
Second, \citet{Ed16} also pointed out that the isotropic Plummer
model has the remarkable property that it obeys the virial theorem
locally (of course, any self-gravitating system must obey it
globally). At every spot, the kinetic energy in each element ($T =
\rho\langle v^2\rangle/2$) is exactly one-half of the magnitude of the local
potential energy ($W = -\rho\psi/2$). Therefore, the virial theorem
holds ($2T + W =0$) at each and every spot! Such models we
call {\it hypervirial}.

In this paper, we introduce a new family of cusped galaxy models. Each
member of the family has a distribution function that rivals
equation~(\ref{eq:plummerdf}) in its simplicity. Even more remarkably, each
member of the family satisfies the virial theorem locally. The Plummer
model is simply the limiting case when there is no cusp.

\section{Intrinsic Properties} 

The potential $\psi$ and density $\rho$ of the family of models are
\begin{equation}
\psi = \frac{GM}{(a^p+r^p)^{1/p}},
\qquad
\rho = \frac{(p+1)\,M}{4\pi}\ \frac{a^p}{r^{2-p}(a^p+r^p)^{2+1/p}}
\label{eq:pdpair}
\end{equation}
where $p$ is a parameter that is a positive real number. The model is
the limiting case of the `generalized isochronous' models first
written down by \citet{Ve79,Ve79tr}, who also worked out some of their
integrated properties. He, however, presented the
potential-density pair (\ref{eq:pdpair}) as a useful fitting formula
and did not study its dynamical basis.

The case $p=2$ is recognized as the Plummer model, while the case
$p=1$ corresponds to a model introduced by \citet{He90}.
Generally, the density distribution behaves like $r^{-(2-p)}$
at small radii and falls off like $r^{-(p+3)}$ at large radii. Note
that, if $p>2$, there is a hole at the centre [$\rho(0)=0$] and the
density increases outwards near the centre. The model with $p=\infty$
is a shell of mass $M$ and radius $a$.  Models with $p>2$ are
therefore not astrophysically realistic. Unless otherwise noted, we
only consider models for which the parameter $p$ is restricted to lie
in the range $0<p\le2$. Henceforth, we use units in which $G=M=a=1$.

The cumulative mass $M_r$ within the sphere of radius of $r$ is
\begin{equation}
M_r 
=4\pi\int_0^r\!\rho r^2\mathrm dr=-r^2\frac{\mathrm d\psi}{\mathrm dr}
=\frac{r^{p+1}}{(1+r^p)^{1+1/p}}
=\frac{1}{(1+r^{-p})^{1+1/p}},
\end{equation}
and so the half-mass radius $r_{1/2}$ is 
\begin{equation}
r_{1/2} = [2^{p/(p+1)}-1]^{-1/p}.
\end{equation}
The circular velocity curve $v_\mathrm{circ}$ is
\begin{equation}
v_\mathrm{circ}^2
=-r\frac{\mathrm d\psi}{\mathrm dr}=\frac{M_r}{r}
=\frac{r^p}{(1+r^p)^{1+1/p}}.
\end{equation}
As $r \rightarrow 0$, the circular velocity tends to zero. As $r
\rightarrow \infty$, the circular velocity becomes asymptotically
Keplerian.  The velocity dispersion is determined by solving the Jeans
equation
\begin{equation}
\frac{1}{\rho}\frac{\mathrm d}{\mathrm dr}\left(\rho\langle v_r^2\rangle\right)
+2\beta\frac{\langle v_r^2\rangle}{r}=\frac{\mathrm d\psi}{\mathrm dr}
\label{eq:Jeans}
\end{equation}
where $\beta$ is the Binney's anisotropy parameter $1-\beta=\langle
v_\mathrm T^2\rangle/(2\langle v_r^2\rangle)$, and $\langle v_r^2 \rangle$ and
$\langle v_\mathrm T^2\rangle$ are the squares of the radial and tangential
velocity dispersions.  In general, for a given potential-density pair,
if the behaviour of $\beta$ is assumed, then the Jeans
equation~(\ref{eq:Jeans}) can be solved using an integrating
factor. In particular, if $\beta$ is a constant, then the radial
velocity dispersion becomes
\begin{eqnarray}
\lefteqn{
\langle v_r^2\rangle
=\frac{1}{r^{2\beta}\rho}\int_\infty^r\!\mathrm dr\
r^{2\beta}\rho\frac{\mathrm d\psi}{\mathrm dr}
}\nonumber\\&&
=\frac{(1+r^p)^{2+1/p}}{r^{p-2+2\beta}}\int_{r}^\infty
\!\frac{r^{2p-3+2\beta}\,\mathrm dr}{(1+r^p)^{3+2/p}}
\nonumber\\&&
=\frac{\psi}{p+4-2\beta}\
\mbox{}_2F_1\!\left(
\frac{2-2\beta}{p}-1,1;\frac{4-2\beta}{p}+2;-\frac{1}{r^p}\right)
\label{eq:vdis}
\end{eqnarray}
where $\mbox{}_2F_1(a,b;c;x)$ is the Gaussian hypergeometric function.
In general, $\langle v_r^2\rangle$ is finite everywhere if
$2\beta\le(2-p)$, while it diverges at the centre if $2\beta>(2-p)$.
In fact, we find that the models with $2\beta>(2-p)$ are unphysical as
the distribution function is not everywhere non-negative
\citep[Appendix~\ref{app}; see also][]{AE05}.

\subsection{Hypervirial Models}

It is of course awkward to work with hypergeometric functions, and so
it is natural to look for simplifications. Fortunately, a wonderful
simplification exists. If $2\beta=(2-p)$ in equation~(\ref{eq:vdis}), then
the hypergeometric function becomes the constant unity! So, the square
of the velocity dispersion becomes linearly proportional to the
potential everywhere, i.e.,
\[\langle v_r^2\rangle
=\frac{1}{2p+2}\frac{1}{(1+r^p)^{1/p}}=\frac{\psi}{2(p+1)},
\]\begin{equation}
\langle v_\mathrm T^2\rangle=2(1-\beta)\langle v_r^2\rangle
=\frac{p\psi}{2(p+1)}.
\end{equation}
Furthermore, we deduce the unusual result that the total velocity
dispersion, $\langle v_r^2\rangle+\langle v_\mathrm T^2\rangle=\psi/2$,
is independent of $p$.  This leads us to suspect that the distribution
functions of these models must also be very simple, though they must
be dependent on the angular momentum $L$ and the binding energy $E$
since the velocity dispersion tensor is in general anisotropic.  It is
straightforward to show that if the angular momentum dependence of the
distribution function is in the form of a power law, i.e.,
$f(E,L)=L^{-2\beta}f_E(E)$, then the velocity dispersion anisotropy is
everywhere constant and the Binney's parameter becomes the constant
$\beta$ \citep[see e.g.,][]{BT87}.  Motivated by this, we
expect the distribution functions of our models will also have the
form of $f(E,L)=L^{p-2}f_E(E)$, where $f_E(E)$ is a function of energy 
that remains to be found.

A little more work shows that $f_E(E)$ is itself a power law. This follows
because the density can be decomposed in terms of a product of powers
of the potential and radius
\begin{equation}
\rho = \frac{p+1}{4\pi}\ r^{p-2}\psi^{2p+1},
\end{equation}
and consequently simple distribution functions exist of the form
\citep[see e.g., eq.~B7 of][]{Ev94}
\begin{equation}
f(E,L) = C L^{p-2} E^{(3p+1)/2}.
\label{eq:hvdfunc}
\end{equation}
Note that $f(E,L=0)$ diverges for $p<2$ which is in accordance with
the presence of the density cusp at the centre.  Here, the constant
$C$ can be determined by the normalization condition
$\rho=\int f\mathrm d^3\bmath v$ and is
\begin{equation}
C = \frac{1}{2^{p/2+1}(2\pi)^{5/2}}
\frac{\Gamma(2p+3)}{\Gamma(p/2)\Gamma(3p/2+3/2)}
\end{equation}
where $\Gamma(x)$ is the Gamma function. The velocity dispersion
tensor is found from the second velocity moments of the distribution
function,
\begin{eqnarray*}\lefteqn{
\rho\langle v_r^2 \rangle=\int\!v_r^2\,f\,\mathrm d^3\!\bmath v
}\\&&
=\frac{2\pi C}{r^{2-p}}
\int_0^\pi\!\mathrm d\theta_v\cos^2\theta_v\sin^{p-1}\!\theta_v
\int_0^{\sqrt{2\psi}}\!\mathrm dv\,v^{p+2}
\left(\psi-\frac{v^2}{2}\right)^{(3p+1)/2}
\\&&
=\frac{1}{8\pi}r^{p-2}\psi^{2p+2},
\end{eqnarray*}
\begin{eqnarray*}\lefteqn{
\rho\langle v_\mathrm T^2 \rangle
=\int\!(v_\theta^2+v_\phi^2)\,f\,\mathrm d^3\!\bmath v
}\\&&
=\frac{2\pi C}{r^{2-p}}\int_0^\pi\!\mathrm d\theta_v\sin^{p+1}\!\theta_v
\int_0^{\sqrt{2\psi}}\!\mathrm dv\,v^{p+2}
\left(\psi-\frac{v^2}{2}\right)^{(3p+1)/2}
\\&&
=\frac{p}{8\pi}r^{p-2}\psi^{2p+2},
\end{eqnarray*}
so that
\[\langle v_r^2 \rangle
=\frac{1}{8\pi}\frac{r^{p-2}\psi^{2p+2}}{\rho}
=\frac{\psi}{2(p+1)},
\]\begin{equation}
\langle v_\mathrm T^2 \rangle
=\frac{p}{8\pi}\frac{r^{p-2}\psi^{2p+2}}{\rho}
=\frac{p\psi}{2(p+1)},
\end{equation}
and
\begin{equation}
\beta =
1-\frac{\langle v_\mathrm T^2\rangle}{2\langle v_r^2\rangle}=1-\frac{p}{2}.
\end{equation}
When $p=2$ (the Plummer model), the velocity dispersion is isotropic
and the distribution function no longer depends on the angular
momentum. As $p \rightarrow 0$, the density of the models become
increasingly cusped and the velocity distribution becomes increasingly
dominated by radial orbits.

Note that the kinetic energy $T$ in any element at each point is
\begin{equation}
T=\frac{\rho}{2}\left(\langle v_r^2\rangle+\langle v_\mathrm T^2\rangle\right)
=\frac{1}{4}\rho\psi=-\frac{W}{2}
\end{equation}
where $W=-(\rho\psi)/2$ is the local contribution to the potential
energy. In other words, we have established that there exists a virial
relation $2T+W=0$ that holds at every spot. All these models are
therefore hypervirial.  This generalizes the remarkable result that
\citet{Ed16} originally found for the Plummer model.

In fact, \citet{BD02} have already noted that the
Hernquist model with $\beta=1/2$ can be be constructed from the
distribution function
\begin{equation}
f(E,L)=\frac{3}{4\pi^3}\frac{E^2}{L}.
\label{eq:herndf}
\end{equation}
This is a particular case of equation~(\ref{eq:hvdfunc}) when $p=1$.
In fact, all the models described by equation~(\ref{eq:hvdfunc})
constitute special cases of the `generalized polytropes'
\citep*[see e.g.,][]{He73,BGH86} with their potential
expressible as elementary functions of the radial distance.  As we
show in the next section, they are also the only generalized
polytropes with a finite mass and an infinite extent.

\subsection{Power-law distribution functions}

Now, let us ask the question: are there any more spherical models with
simple distribution functions that are just a power of energy
multiplied by a power of angular momentum?

Suppose that the distribution function of a spherically symmetric
system is indeed given by the ansatz
\begin{equation}
f(E,L)=C L^{-2\beta} E^{n-3/2}
\label{eq:simpledf}
\end{equation}
where $\beta<1$ and $n>1/2$. Then, the corresponding density becomes
\begin{eqnarray}
\rho=\int f\mathrm d^3\!\bmath v
=2^{3/2-\beta}\pi^{3/2}C\,
\frac{\Gamma(1-\beta)\Gamma(n-1/2)}{\Gamma(n-\beta+1)}\
\frac{\psi^{n-\beta}}{r^{2\beta}}.
\end{eqnarray}
The velocity dispersions may also be found as
\begin{eqnarray*}\lefteqn{
\langle v_r^2\rangle=\frac{1}{\rho}\int v_r^2 f\mathrm d^3\!\bmath v
=2^{3/2-\beta}\pi^{3/2}C\,
\frac{\Gamma(1-\beta)\Gamma(n-1/2)}{\Gamma(n-\beta+2)}\
\frac{\psi^{n-\beta+1}}{\rho r^{2\beta}}
}\\&&
=\frac{\psi}{n-\beta+1},
\end{eqnarray*}
\begin{eqnarray}\lefteqn{
\langle v_\mathrm T^2\rangle=\frac{1}{\rho}\int(v_\theta^2\!+\!v_\phi^2)
f\mathrm d^3\!\bmath v=2^{5/2-\beta}\pi^{3/2}C\,
\frac{\Gamma(2\!-\!\beta)\Gamma(n\!-\!1/2)}{\Gamma(n\!-\!\beta+2)}\
\frac{\psi^{n-\beta+1}}{\rho r^{2\beta}}
}\nonumber\\&&
=\frac{2(1-\beta)\psi}{n-\beta+1}.
\end{eqnarray}
However, the total kinetic energy $T_\mathrm{tot}$ is
\begin{eqnarray}\lefteqn{
T_\mathrm{tot}=
\frac{1}{2}\int\!\rho(\langle v_r^2\rangle
+\langle v_\mathrm T^2\rangle)\,\mathrm d^3\!\bmath r
=\frac{1}{2}\int\!\mathrm d^3\!\bmath r\ \rho\ \frac{(3-2\beta)\psi}{n-\beta+1}
}\nonumber\\&&
=-\frac{3-2\beta}{n-\beta+1}W_\mathrm{tot}
\end{eqnarray}
where $W_\mathrm{tot}$ is the total potential energy.
In other words, the sum $W_\mathrm{tot}+2T_\mathrm{tot}$ is
\begin{equation}
W_\mathrm{tot}+2T_\mathrm{tot}
=\left[1-\frac{2(3-2\beta)}{n-\beta+1}\right]W_\mathrm{tot}
=\frac{n+3\beta-5}{n-\beta+1}W_\mathrm{tot},
\end{equation}
suggesting that the global virial theorem for a steady-state system is
satisfied only if $n+3\beta=5$ [or, in other words, if
$\beta=1-(p/2)$, then $n=2+(3/2)p$ and $n-3/2=(3p+1)/2$]. In this
case, the total potential energy can be explicitly evaluated as
\begin{equation}
W_\mathrm{tot} = -2T_\mathrm{tot} 
= -2\pi \int_0^\infty\!\rho\,\psi\,r^2\,\mathrm dr
= -\frac{\pi^{1/2}}{2^{2/p+2}}\frac{\Gamma(1/p+2)}{\Gamma(1/p+3/2)} 
\end{equation}
where $p=2(1-\beta)=2(n-2)/3$.  So, our hypervirial models are the
only spherically symmetric steady-state systems with finite mass which
have distribution functions as simple as
equation~(\ref{eq:simpledf}). In particular, the density profile of
the Plummer model is the only spherically symmetric self-gravitating
polytrope of finite mass with an infinite extent that can be thermally
(pressure) supported.

There are other known spherical models which do have distribution
functions as simple as equation~(\ref{eq:simpledf}) -- namely, the
power-law spheres \citep{Ev94}. However, these models have infinite
mass and so do not satisfy the global virial theorem (unless boundary
terms are added). Similarly, stellar dynamical polytropes also exist,
but have distribution functions $f \propto (E-E_0)^N$, where $E_0$ and
$N$ are constants \citep{Ed16}. Only for the Plummer model does
$E_0$ vanish and so the model is of infinite extent.

\section{Projected Quantities}

The beauty of the hypervirial models lies in their simple
distributions of velocities. In contrast, the projected quantities are
generally more awkward, typically reducing to elementary functions
only in the cases of Plummer ($p=2$) and Hernquist ($p=1$).\footnote{
Although we do not derive the explicit forms, the quantities for
the case of $p=1/2$ in general reduces to the expression involving
elliptic integrals.}

\begin{figure}
\includegraphics[width=\hsize]{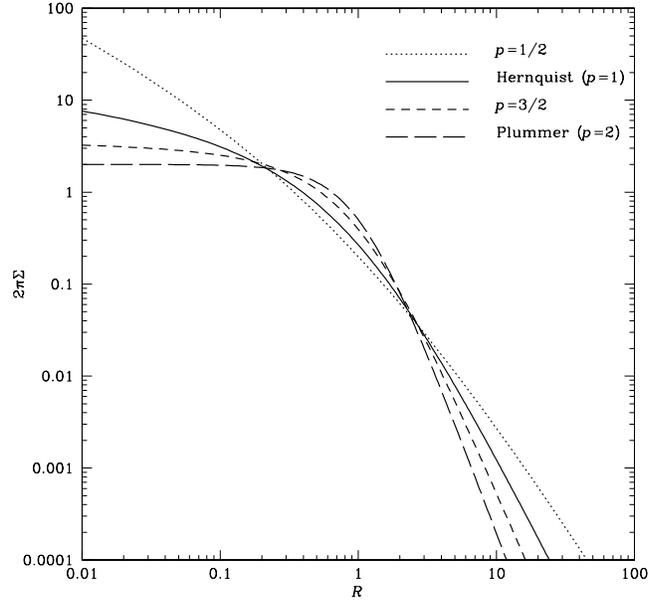}
\caption{\label{fig:sdens} The surface density profile of the
hypervirial models with $p=1/2, 1$ (Hernquist), $3/2$ and $2$
(Plummer).}
\end{figure}

The surface density is
\begin{equation}
\Sigma=2\int_R^\infty\!\frac{\rho r\,\mathrm dr}{\sqrt{r^2-R^2}}
=\frac{p+1}{2\pi}\,R^{p-1}\!\int_0^{\pi/2}\!
\frac{\cos^{p+1}\!\theta\,\mathrm d\theta}{(R^p+\cos^p\!\theta)^{2+1/p}}.
\end{equation}
If the mass-to-light ratio is constant, then this is proportional to
the surface brightness. Fig.~\ref{fig:sdens} shows the surface
brightness profile for a number of the hypervirial models.  For the
two special cases, expressions using elementary functions are known
\citep{Pl11,He90}
\[\Sigma_\mathrm P=\frac{1}{\pi}\frac{1}{(1+R^2)^2}
\qquad\mbox{Plummer}\ (p=2),
\]\[
\Sigma_\mathrm H=\frac{1}{2\pi}\frac{1}{(1-R^2)^2}
\left[(2+R^2)\,X(R)-3\right]
\qquad\mbox{Hernquist}\ (p=1),
\]
where, the Hernquist function $X(R)$ is defined as
\[
X(R)\equiv\frac{2}{1+R}\,\mbox{}_2F_1\!
\left(1,\frac{1}{2};\frac{3}{2};\frac{1-R}{1+R}\right)
=\frac{\mbox{arcsech }R}{\sqrt{1-R^2}}
=\frac{\mbox{arcsec }R}{\sqrt{R^2-1}}.
\]
Despite the appearance of the formal singularity at $R=1$ for the
$p=1$ case, $\Sigma_\mathrm H$ is in fact regular everywhere for $R>0$. In
particular, it is continuous [$\Sigma_\mathrm H(1)=(2\pi)^{-1}(4/15)$] and
differentiable at $R=1$.

The central surface density is finite if $p>1$,
\begin{equation}
\Sigma(0)=2\int_0^\infty\!\rho\,\mathrm dr
=\frac{\Gamma(1-1/p)\Gamma(1+2/p)}{2\pi\Gamma(1+1/p)},
\end{equation}
but it is divergent otherwise. In particular, if $0<p<1$, the
central surface density behaves like
\begin{equation}
\Sigma\sim
\frac{1}{R^{1-p}}\frac{(p+1)\Gamma(1/2-p/2)}{4\pi^{1/2}\Gamma(1-p/2)}
\propto R^{-(1-p)}
\qquad(R\rightarrow0),
\end{equation}
and it diverges logarithmically if $p=1$ \citep{He90}.  At large
radii, the asymptotic behaviour of the surface density is given by
\begin{equation}
\Sigma \sim \frac{1}{R^{p+2}}
\frac{\Gamma(1+p/2)}{2\pi^{1/2}\Gamma(1/2+p/2)}
\propto R^{-(p+2)}
\qquad(R\rightarrow\infty).
\end{equation}

\begin{figure}
\includegraphics[width=\hsize]{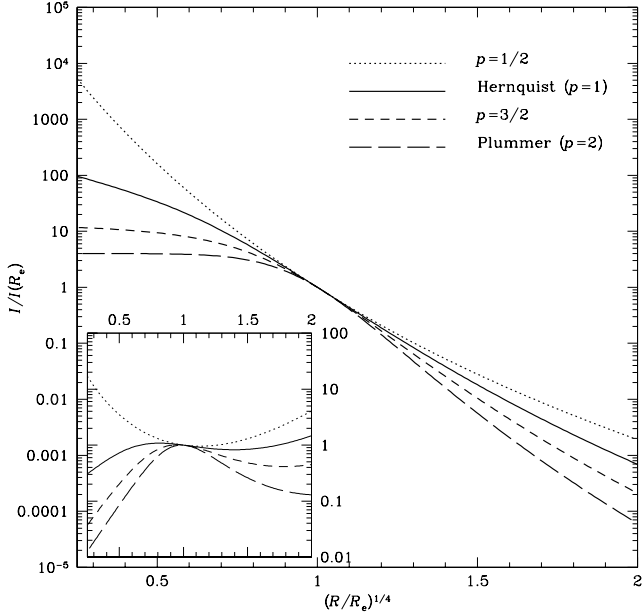}
\caption{\label{fig:rquarter} The surface brightness of the
hypervirial models with $p=1/2, 1$ (Hernquist), $3/2$ and $2$
(Plummer) plotted against $(R/R_\mathrm e)^{1/4}$ where
$R/R_\mathrm e$ is the projected distance in units of the effective
radius of each model. In this diagram, the \citet{dV48}
profile would be a straight line with a slope of $-3.331$.  The inset
shows the residuals with respect to the de Vaucouleurs profile.}
\end{figure}

If the mass-to-light ratio is constant, then the cumulative brightness
is proportional to the mass within the cylinder of radius of $R$,
namely
\begin{equation}
\mathcal{M}_R=2\pi\int_0^R\!\Sigma R\,\mathrm dR =R^{p+1}\int_0^{\pi/2}\!
\frac{\cos\theta\,\mathrm d\theta}{(R^p+\cos^p\!\theta)^{1+1/p}}.
\end{equation}
For the Plummer and Hernquist models, the result reduces to
\[\mathcal{M}_R^\mathrm P=\frac{R^2}{1+R^2}
\qquad\mbox{Plummer}\ (p=2),
\]\[
\mathcal{M}_R^\mathrm H=\frac{R^2}{1-R^2}\left[X(R)-1\right]
\qquad\mbox{Hernquist}\ (p=1).\]
Note that $\mathcal{M}_R^\mathrm P(1)=1/2$ and $\mathcal{M}_R^\mathrm H(1)=1/3$.
The effective radius (or the half-light radius) $R_\mathrm e$ can,
in general, be found by numerically solving
\[
\mathcal{M}_R(R_\mathrm e)=\frac{1}{2}.
\]
The particular solutions are $R_\mathrm e\approx$ 1 (exact), 1.18084,
1.81527 and 11.0151 for $p=$ 2, 3/2, 1 and 1/2, respectively.
Fig.~\ref{fig:rquarter} shows the surface brightness normalized by
$\Sigma(R_\mathrm e)$ as a function of $(R/R_\mathrm e)^{1/4}$. We find that
Hernquist profile -- and by extension the hypervirial models with $p
\approx 1$ -- are all reasonably good approximations to the empirical
de Vaucouleurs $R^{1/4}$ law (\citeyear{dV48}) of the surface brightness
profiles between $0.25\la(R/R_\mathrm e)\la10$. The total mass is of course
\[\mathcal{M}_R(\infty)
=\lim_{R\rightarrow\infty}\int_0^{\pi/2}\!
\frac{\cos\theta\,\mathrm d\theta}{(1+R^{-p}\cos^p\!\theta)^{1+1/p}}
=\int_0^{\pi/2}\!\cos\theta\,\mathrm d\theta=1,\]
as it should be!

\begin{figure}
\includegraphics[width=\hsize]{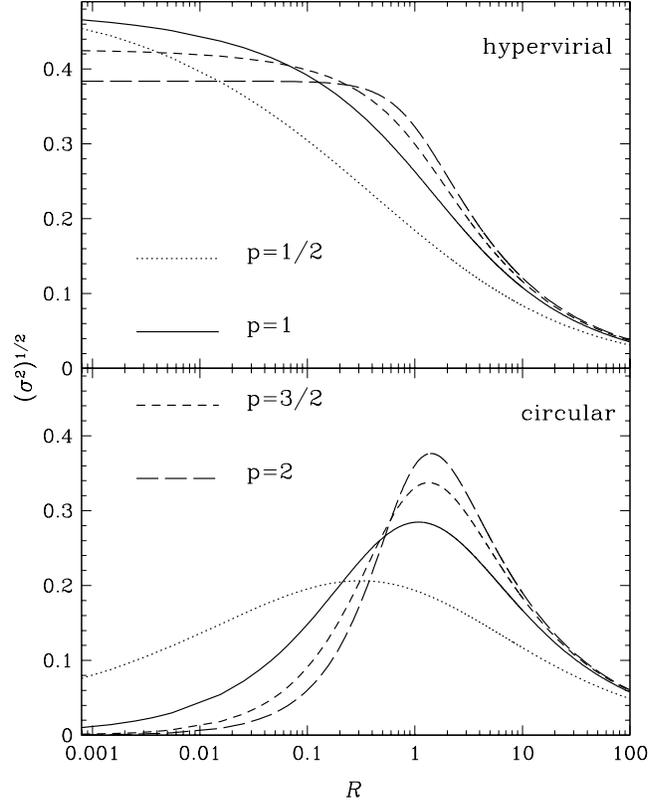}
\caption{\label{fig:vdis} The line-of-sight velocity dispersion
profiles for the hypervirial and circular orbit distribution functions
corresponding to the models with $p=1/2, 1$ (Hernquist), $3/2$ and $2$
(Plummer).}
\end{figure}

For any spherical model with anisotropy parameter $\beta$, the
line-of-sight velocity dispersion $\sqrt{\sigma^2}$ is
\citep[e.g.,][]{BT87}
\begin{equation}
\Sigma\sigma^2=2\int_R^\infty\!\left(1-\beta\frac{R^2}{r^2}\right)
\frac{\rho\langle v_r^2\rangle r\,\mathrm dr}{\sqrt{r^2-R^2}}.
\end{equation}
For the hypervirial models, $\beta=1-(p/2)$, and so
\begin{equation}
\Sigma\sigma^2
=\frac{1}{4\pi}R^{p-1}\int_0^{\pi/2}\!
\frac{\cos^{p+2}\!\theta+(p/2-1)\cos^{p+4}\!\theta}
{(R^p+\cos^p\!\theta)^{2+2/p}}\
\mathrm d\theta.
\end{equation}
As in the case of the surface density, this integral reduces to
elementary functions for the Plummer model
\[\Sigma_\mathrm P\sigma_\mathrm P^2=\frac{3}{64}\frac{1}{(1+R^2)^{5/2}};
\qquad
\sigma_\mathrm P^2=\frac{3\pi}{64}\frac{1}{(1+R^2)^{1/2}},
\]
and for the Hernquist model
\begin{eqnarray}
\lefteqn{
\Sigma_\mathrm H\sigma_\mathrm H^2=\frac{1}{4\pi}
\left\{\pi R + \frac{1}{12(1-R^2)^3}\left[
(24R^6-68R^4+57R^2-28)
\right.\right.}\nonumber\\&&\left.\left.
+3(8R^8-28R^6+35R^4-14R^2+4)\,X(R)\right]\right\}.
\label{eq:hernlos}
\end{eqnarray}
This is continuous at $R=1$ with
$\sigma^2_\mathrm H(1)=[4\pi\Sigma_\mathrm H(1)]^{-1}[\pi-(326/105)]$. Note that,
the isotropic model was the main object of study of \citet{He90}
and its line-of-sight velocity dispersion \citep[eq.~41 of][]{He90}
differs from equation~(\ref{eq:hernlos}), of course.

If $p>1$, then the central value of the velocity dispersion is
\begin{equation}
\sigma^2(0)=
\frac{2}{\Sigma(0)}\int_0^\infty\!\rho\langle v_r^2\rangle\,\mathrm dr
=\frac{\Gamma(1+3/p)\Gamma(1+1/p)}{4\Gamma(2/p)\Gamma(2+2/p)},
\label{eq:centre}
\end{equation}
which varies from $3\pi/64$ for $p=2$ to $1/4$ as $p\rightarrow1^+$.
On the other hand, if $0<p<1$, the central velocity dispersion is
\begin{equation}
\Sigma\sigma^2
\sim\frac{1}{R^{1-p}}\frac{(p+1)\Gamma(1/2-p/2)}{16\pi^{1/2}\Gamma(1-p/2)}
\sim\frac{\Sigma}{4}
\qquad(R\rightarrow0).
\end{equation}
For the case $p=1$, using $X(R)\sim\ln(2/R)$ as
$R\rightarrow0$, we find that
\[\Sigma_\mathrm H\sigma^2_\mathrm H\sim
\frac{1}{\pi}\left(\frac{1}{4}\ln\frac{2}{R}-\frac{7}{12}\right);
\qquad
\Sigma_\mathrm H\sim
\frac{1}{\pi}\left(\ln\frac{2}{R}-\frac{3}{2}\right),
\]\[
\lim_{R\rightarrow0}\sigma^2_\mathrm H
=\lim_{R\rightarrow0}\frac{\Sigma_\mathrm H\sigma^2_\mathrm H}{\Sigma_\mathrm H}
=\frac{1}{4}.
\]
In other word, the line-of-sight velocity dispersion at the centre is
always finite for $0<p\le2$. In particular, $\sigma^2(0)=1/4$ if
$0<p\le1$ while it can be determined from equation~(\ref{eq:centre})
for $1\le p\le2$.  At large radii, the fall-off becomes Keplerian
\begin{equation}
\sigma^2
\sim\frac{1}{R}\frac{(p+2)\Gamma(3/2+p/2)^2}{8\Gamma(3+p/2)\Gamma(1+p/2)}
\propto R^{-1}
\qquad(R\rightarrow\infty).
\label{eq:asym}
\end{equation}

\begin{figure}
\includegraphics[width=\hsize]{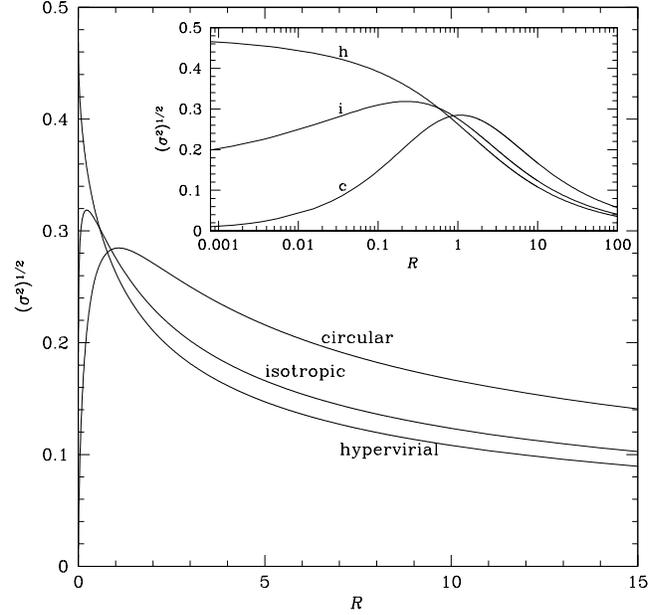}
\caption{\label{fig:vhern} The line-of-sight velocity dispersion
profiles for the isotropic, hypervirial and circular orbit
distribution functions corresponding to the Hernquist model
\citep[c.f., fig.~1 of][]{He90}. The inset is the same but as a function of
$\log R$, which shows the behaviour near the centre better.}
\end{figure}

It is instructive to compare this with the purely circular orbit
model, for which the line-of-sight velocity dispersion is given by
\begin{eqnarray}
\lefteqn{
\Sigma\sigma^2=
2\int_R^\infty\!\frac{\langle v_\mathrm T^2\rangle}{2}\frac{R^2}{r^2}
\frac{\rho r\,\mathrm dr}{\sqrt{r^2-R^2}}
=R^2\int_R^\infty\!\frac{\rho v_\mathrm{circ}^2\,\mathrm dr}{r\sqrt{r^2-R^2}}
}\nonumber\\&&
=\frac{p+1}{4\pi}R^{2p-1}\int_0^{\pi/2}\!
\frac{\cos^{p+4}\!\theta\,\mathrm d\theta}{(R^p+\cos^p\!\theta)^{3+2/p}},
\end{eqnarray}
which becomes
\[\Sigma_\mathrm P\sigma_\mathrm P^2=\frac{15}{128}\frac{R^2}{(1+R^2)^{7/2}};
\qquad
\sigma_\mathrm P^2=\frac{15\pi}{128}\frac{R^2}{(1+R^2)^{3/2}}
\qquad(p=2),
\]
\begin{eqnarray*}
\lefteqn{
\Sigma_\mathrm H\sigma_\mathrm H^2=\frac{1}{4\pi}
\left\{\pi R - \frac{R^2}{12(1-R^2)^4}\left[
(24R^6-92R^4+117R^2-154)
\right.\right.}\\&&\left.\left.
+3(8R^8-36R^6+63R^4-40R^2+40)\,X(R)\right]\right\}
\qquad(p=1).
\end{eqnarray*}
Since we only observe radially directed motion when looking at the
very centre, the line-of-sight velocity dispersion must vanish there
for the extreme tangentially anisotropic model. On the other hand, at
large radii, we have
\begin{equation}
\sigma^2
\sim\frac{1}{R}
\frac{\Gamma(5/2+p/2)\Gamma(3/2+p/2)}{2\Gamma(3+p/2)\Gamma(1+p/2)}
\propto R^{-1}
\qquad(R\rightarrow\infty).
\end{equation}
By comparing this to equation~(\ref{eq:asym}), we find that the
line-of-sight velocity dispersion for the purely circular orbit model is
larger than that for the hypervirial model at large radii. This is an
expected result, since we observe tangential motion preferentially as
the lines of sight moves to the outskirts of the system.  In
Fig.~\ref{fig:vdis}, we plot the line-of-sight velocity dispersion for
the hypervirial and circular orbit model for a few representative
values of $p$. In Fig.~\ref{fig:vhern}, the line-of-sight velocity
dispersions for the Hernquist model are shown. The expression for the
velocity dispersion of the isotropic model is provided by
equation~(41) of \citet{He90}.

\section{Hyperviriality}

Next, let us ask the question: are there any more spherical models
with the property of hyperviriality? Mathematically, such a model
has to satisfy
\begin{equation}
\label{eq:hvirial}
\langle v_r^2\rangle+\langle v_\mathrm T^2\rangle=\frac{\psi}{2}.
\end{equation}
Since $1-\beta=\langle v_\mathrm T^2\rangle/(2\langle v_r^2\rangle)$,
if $\beta$ is constant, equation~(\ref{eq:hvirial}) implies that
\begin{equation}
\label{eq:prop}
\langle v_r^2\rangle=\frac{\psi}{2(3-2\beta)}.
\end{equation}
On the other hand, the Jeans equation~(\ref{eq:Jeans}) can be written
\begin{equation}
\label{eq:J2}
\frac{\mathrm d\langle v_r^2\rangle}{\mathrm dr}
+\left(\frac{1}{\rho}\frac{\mathrm d\rho}{\mathrm dr}+\frac{2\beta}{r}\right)
\langle v_r^2\rangle=\frac{\mathrm d\psi}{\mathrm dr}.
\end{equation}
From equations~(\ref{eq:prop}) and (\ref{eq:J2}), by eliminating
$\langle v_r^2\rangle$, we obtain
\begin{equation}
\label{eq:J3}
\frac{\mathrm d\rho}{\mathrm dr}=
\left[\left(5-4\beta\right)\frac{\mathrm d\ln\psi}{\mathrm d\ln r}-2\beta\right]
\frac{\rho}{r}.
\end{equation}
Now, Poisson's equation for a spherically symmetric system reads
\begin{equation}
\label{eq:P}
\frac{1}{r^2}\frac{\mathrm d}{\mathrm dr}
\left(r^2\frac{\mathrm d\psi}{\mathrm dr}\right)=-4\pi G\rho
\end{equation}
where the negative sign is due to our choice of the sign for the
potential.  By eliminating $\rho$ from equations~(\ref{eq:J3}) and
(\ref{eq:P}), we can get a third-order non-linear differential equation
for $\psi$. However, using the relation $\mathrm d\psi/\mathrm
dr=\psi(\mathrm d\ln\psi/\mathrm dr)$, it is possible to reduce the
equation into a second-order differential equation for $\chi=(\mathrm
d\ln\psi/\mathrm dr)$,
\begin{eqnarray}\lefteqn{
\frac{\mathrm d^2\chi}{\mathrm dr^2}+2\frac{\mathrm d\chi}{\mathrm dr}\left[
\frac{1+\beta}{r}-(1-2\beta)\chi\right]
}\nonumber\\&&
-2\left[(1-2\beta)\frac{\chi}{r^2}+(4-5\beta)\frac{\chi^2}{r}+
2(1-\beta)\chi^3\right]=0,
\end{eqnarray}
or additionally changing the independent variable into $t=\ln r$,
\begin{equation}
\label{sec}
\frac{\mathrm d^2\xi}{\mathrm dt^2}-(1-2\beta)(2\xi+1)
\frac{\mathrm d\xi}{\mathrm dt}
-2(1-\beta)(2\xi+1)(\xi+1)\xi=0
\end{equation}
where $\xi=(\mathrm d\ln\psi/\mathrm d\ln r)=(\mathrm d\ln\psi/\mathrm dt)$.

Since equation~(\ref{sec}) does not involve the independent variable, its
order may be reduced by the substitution \citep[e.g,][]{In44}
\[
\zeta=\frac{\mathrm d\xi}{\mathrm dt};
\qquad
\frac{\mathrm d^2\xi}{\mathrm dt^2}
=\frac{\mathrm d\zeta}{\mathrm dt}
=\frac{\mathrm d\zeta}{\mathrm d\xi}\frac{\mathrm d\xi}{\mathrm dt}
=\zeta\frac{\mathrm d\zeta}{\mathrm d\xi},
\]
so that one finally arrives at
\begin{equation}
\zeta\frac{\mathrm d\zeta}{\mathrm d\xi}-(1-2\beta)(2\xi+1)\zeta
-2(1-\beta)(2\xi+1)(\xi+1)\xi=0.
\end{equation}
Since this is a first-order equation, it is always possible (at least
formally) to find an integrating factor, which in this case is
\[
I=\frac{1}{[\zeta+\xi(\xi+1)][\zeta-2(1-\beta)\xi(\xi+1)]}.
\]
Then, the equation can be written in the exact form
\[
\frac{\mathrm d}{\mathrm d\xi}
\ln\left|\left[\zeta-2(1-\beta)\xi(\xi+1)\right]^{2(1-\beta)}
\left[\zeta+\xi(\xi+1)\right]\right|=0.
\]
If $\beta<1$, we obtain the solution
\begin{equation}
\left[\zeta-2(1-\beta)\xi(\xi+1)\right]^{2(1-\beta)}
\left[\zeta+\xi(\xi+1)\right]=C
\label{eq:cdef}
\end{equation}
where $C$ is constant. If we restrict attention to systems of finite
mass, then the potential is asymptotically $\sim r^{-1}$ and consequently
we can set $C=0$ from the boundary condition at $r=\infty$
($\xi=-1$ and $\zeta=0$). Then, we find two possible solutions, namely,
\[
\zeta=2(1-\beta)\xi(\xi+1),
\qquad
\zeta=-\xi(\xi+1).
\]
Note that the second solution is independent of $\beta$. In fact, we
find that it leads to vanishing density everywhere so that the second
solution is unphysical.  For the first solution, we obtain
\[
\zeta=\frac{\mathrm d\xi}{\mathrm dt}=p\xi(\xi+1),
\qquad
\frac{\mathrm d\xi}{\xi(\xi+1)}=p\,\mathrm dt
\]
where $p=2(1-\beta)$. Then, integrating gives the result that
\begin{equation}
\label{eq:xi}
\ln\left|\frac{\xi}{\xi+1}\right|=p(t-t_0),
\qquad
\xi=-\frac{1}{1\pm\mathrm e^{-p(t-t_0)}}
\end{equation}
where the double sign in front of the exponential appears when the
absolute value is removed.  However, since $\xi=(\mathrm
d\ln\psi/\mathrm dt)$, one can integrate equation~(\ref{eq:xi})
further to find $\psi$, that is,
\begin{equation}
\ln\left|\frac{\psi}{\psi_0}\right|
=-\int\frac{\mathrm dt}{1\pm\mathrm e^{-p(t-t_0)}}
=-\frac{1}{p}\ln\left|1\pm\mathrm e^{p(t-t_0)}\right|,
\label{eq:absval}
\end{equation}
so that
\begin{equation}
\psi=\psi_0\left[1\pm\mathrm e^{p(t-t_0)}\right]^{-1/p}
=\frac{\psi_0}{[1\pm (r/r_0)^p]^{1/p}}.
\label{eq:soln}
\end{equation}
Note that the absolute value in equation~(\ref{eq:absval}) can be
ignored since $\psi_0$ can be either negative or positive. The
solution corresponding to the negative sign in
equation~(\ref{eq:soln}) is unphysical, while the other solution
recovers our hypervirial models.  One integration constant ($t_0=\ln
r_0$) corresponds to the scalelength and the other integration
constant ($\psi_0$) becomes the overall scaling factor.

Our original differential equation before reduction was in third
order, so we expect the general solution to contain three constants of
integration. The hypervirial family is a two-parameter family, and so
there exists an additional family of solutions, corresponding to the
choice $C\neq0$ in equation~(\ref{eq:cdef}) and consequently having
either infinite mass or finite extent. Unfortunately, the
potential-density pair now seems not to be expressible in terms of
elementary functions.

Physically speaking, hyperviriality is related to stability of the
model against evaporation. The escape speed is $\sqrt{2\psi}$. In a
hypervirial model, the root mean square speed is $\sqrt{\psi/2}$. For
any other kind of model, the root mean square speed may lie below the
hypervirial value at some places, provided that there is suitable
compensation at other places so that the global virial theorem is
obeyed. In other words, at some spots, there will be more stars close
to the escape speed and hence on the verge of escaping. If a few stars
escape by accident or tidal perturbations, then the potential is
lowered and stars originally safe would be left with speeds above the
escape speed. Hyperviriality therefore aids stability by minimizing
the number of high-velocity stars.

However, evaporation is not the only cause of instability. Dynamical
effects, such as the radial orbit instability, may drive spherical
stellar systems into triaxiality. In particular, \citet{PP87}
argue that any distribution functions unbounded at zero angular
momentum, as are all those given by equation~(\ref{eq:hvdfunc}) except
for $p=2$, are formally unstable, though the growth rate of any
instability may be slow. Simulations of generalized polytropes have
already been carried out by \citet{He73} and \citet{BGH86}.
Such numerical work has tried
to identify a critical ratio of kinetic energy in radial to tangential
motion $2T_r/T_\mathrm{T}$ above which the radial orbit instability
occurs. This quantity is $2/p$ for the hypervirial models and so is 2
for the Hernquist model. The stability criterion $2T_r/T_\mathrm{T}<2$
has been suggested as a crude summary of a wide range of numerical
experiments \citet{BGH86}. Probably, only full-scale numerical
simulations of the hypervirial models can find the exact point at
which the radial orbit instability sets in, but it seems reasonable to
expect the most highly cusped members (which are the most radially
anisotropic) to be susceptible to the radial orbit instability.

\section{Conclusions}

This paper has provided a set of very simple distribution functions
for a family of cusped spherical galaxy models. If the density is
cusped like $\rho \sim r^{p-2}$ at small radii, then there is a simple
anisotropic distribution function, which behaves like $f \propto
L^{p-2} E^{(3p+1)/2}$ (where $E$ is the binding energy and $L$ is the
angular momentum). We call these models the hypervirial family. This
is because every model obeys the virial theorem at each and every
spot, and, of course globally.

The family includes the Hernquist model which possesses the
cosmologically important $r^{-1}$ cusp at small radii and has a simple
distribution function $f \propto L^{-1} E^2$. As its sole isotropic
and uncusped representative, the family includes the Plummer model
with its familiar distribution function $f \propto E^{7/2}$.  There
are also members which possess other density cusps such as $\rho \sim
r^{-4/3}$ and $\rho \sim r^{-3/2}$, which have been suggested as
important on cosmogonic grounds \citep{EC97,Mo98}.

In a sense, the models in this paper really are the last word in
simplicity -- for we have proved that they are the only
self-gravitating, spherically symmetric finite mass models with such
simple distribution functions.

\bibliography{../reference}

\appendix
\section{Constant-$\bbeta$ Distribution Function}
\label{app}

Systems with a constant velocity anisotropy parameter always can be built
from the distribution functions of the form
\[
f(E,L)=L^{-2\beta}f_E(E)
\]
where $f_E(E)$ is a function of the binding energy alone
\citep[e.g.,][]{BT87}. This is a particular case of the distribution
function investigated by \citet{Cu91} at the limit of his parameter
$r_a\rightarrow\infty$, and thus it is straightforward to derive an inversion
for the unknown function $f_E(E)$, which is scarcely any more difficult
than Eddington's inversion (eq.~\ref{eq:eddington})
\citep[e.g., eq.~24 of][see also \citealt{De86,WE99,BD02}]{Cu91};
\begin{eqnarray}\lefteqn{
f(E,L)=\frac{1}{L^{2\beta}}
\frac{2^\beta}{(2\pi)^{3/2}}\frac{1}{\Gamma(1-\alpha)\Gamma(1-\beta)}\
\frac{\mathrm d}{\mathrm dE}\int_0^E\frac{\mathrm d\psi}{(E-\psi)^\alpha}
\frac{\mathrm d^nh}{\mathrm d\psi^n}
}\nonumber\\&&
=\frac{2^\beta}{(2\pi)^{3/2}}
\frac{1}{\Gamma(1-\alpha)\Gamma(1-\beta)}
\frac{1}{L^{2\beta}}
\left[\int_0^E\frac{\mathrm d\psi}{(E-\psi)^\alpha}
\frac{\mathrm d^{n+1}h}{\mathrm d\psi^{n+1}}
+\frac{1}{E^\alpha}
\left.\frac{\mathrm d^nh}{\mathrm d\psi^n}\right|_{\psi=0}\right].
\qquad
\label{eq:disint}
\end{eqnarray}
Here, $h(\psi)=r^{2\beta}\rho$ and
$\beta=(3/2)-n-\alpha$ where $n$ is an integer and $0\le\alpha<1$.
Note that, formally, the choice of an arbitrary non-negative integer
for $n$ in equation~(\ref{eq:disint}) actually
does produce the equivalent result. However, strictly speaking,
for $\alpha\ge1$, the integral in general diverges and the formula therefore
becomes meaningless although ad-hoc extension of the formal definitions
can be employed to provide the proper final result.

For the potential-density pair of equation~(\ref{eq:pdpair}),
we find ($G=M=a=1$)
\[
h(\psi)=r^{2\beta}\rho
=\frac{p+1}{4\pi}\psi^{p+3-2\beta}(1-\psi^p)^{1-2(1-\beta)/p}.
\]
Next, using
\begin{equation}
\frac{\mathrm d}{\mathrm d\psi}\left[\frac{\psi^A}{(1-\psi^p)^B}\right]
=A\frac{\psi^{A-1}}{(1-\psi^p)^B}
+pB\frac{\psi^{A-1+p}}{(1-\psi^p)^{B+1}},
\label{eq:dif}
\end{equation}
one finds that
\[
\left.\frac{\mathrm d^mh}{\mathrm d\psi^m}\right|_{\psi=0}=0
\]
for $m<p+3-2\beta$. Then, equation~(\ref{eq:disint}) reduces to
\[
f(E,L)=
\frac{2^{\beta-1}}{(2\pi)^{5/2}}\frac{p+1}{\Gamma(1-\alpha)\Gamma(1-\beta)}\,
\frac{1}{L^{2\beta}}
\int_0^E\!\frac{\mathrm d\psi}{(E-\psi)^\alpha}
\frac{\mathrm d^{n+1}}{\mathrm d\psi^{n+1}}
\left[\frac{\psi^{p+3-2\beta}}{(1-\psi^p)^\gamma}\right]
\]
where $\gamma+1=2(1-\beta)/p\ge0$. If $\gamma\ge0$,
equation~(\ref{eq:dif}) further implies that,
for $0\le E\le\psi\le1$, the integrand is always non-negative
and so is the distribution function.
In fact, it is, at least formally, possible to derive the series
expression of the distribution function for arbitrary $p$ and $\beta$
from the integral form, i.e.,
\[
f(E,L)=
\frac{2^{\beta-1}}{(2\pi)^{5/2}}\frac{p+1}{\Gamma(1-\beta)}\,
\frac{E^{p+3/2-\beta}}{L^{2\beta}}\,\sum_{k=0}^\infty
\frac{\Gamma(pk+p+4-2\beta)}{\Gamma(pk+p+5/2-\beta)}
\frac{\Gamma(k+\gamma)}{\Gamma(\gamma)}
\frac{E^{pk}}{k!}.
\]

On the other hand,
if $-1\le\gamma<0$, equation~(\ref{eq:dif}) also indicates that
\[
\left.\frac{\mathrm d^mh}{\mathrm d\psi^m}\right|_{\psi=1}<0,
\]
and subsequently, one can establish that
\[
\lim_{E\rightarrow1^-}f(E,L)<0
\]
In other words, there exists a certain value $\underline{E}$ such that
$f(E,L)<0$ for $0<\underline{E}<E\le1$, that is, the corresponding
distribution function is unphysical. Hence,
for the potential-density pair of equation~(\ref{eq:pdpair}),
the constant anisotropy distribution function is physical only if
$p\le2(1-\beta)$ and the hypervirial models
which have an anisotropy parameter $\beta=1-(p/2)$ are the models
with the maximally radially biased velocity dispersions
for a given $p$ and a constant $\beta$.

\label{finish}
\end{document}